\documentclass{ifacconf}
\usepackage{amsmath,amssymb,amsfonts}
\usepackage{algorithmic}
\usepackage{graphicx}
\usepackage{textcomp}
\usepackage{xcolor}
\usepackage{lipsum}% http://ctan.org/pkg/lipsum
\usepackage{float}
\usepackage{multirow}
\usepackage{graphicx}      % include this line if your document contains figures
\usepackage{layouts}
\usepackage[T1]{fontenc}
\newcommand\thefont{\expandafter\string\the\font}
\usepackage{color,soul}
\usepackage{xcolor}
\usepackage{soul}
\usepackage{dsfont}
\usepackage{filecontents}

\makeatletter
\let\old@ssect\@ssect % Store how ifacconf defines \@ssect
\makeatother

\usepackage{natbib}
\usepackage{hyperref}

\makeatletter
\def\@ssect#1#2#3#4#5#6{%
  \NR@gettitle{#6}% Insert key \nameref title grab
  \old@ssect{#1}{#2}{#3}{#4}{#5}{#6}% Restore ifacconf's \@ssect
}
\makeatother
\usepackage{lipsum}
\usepackage{titlesec}
\titlespacing\section{0pt}{1pt plus 1pt minus 1pt}{1pt plus 1pt minus 1pt}
\titlespacing\subsection{0pt}{1pt plus 1pt minus 1pt}{1pt plus 1pt minus 1pt}
\newtheorem{definition}{Definition}[section]

\begin{document}
\begin{frontmatter}

\title{Robust tube-based LPV-MPC for autonomous lane keeping\thanksref{footnoteinfo}} 

\thanks[footnoteinfo]{H. S. Abbas is funded by the German Research Foundation (DFG), project number 419290163.
 N.T. Nguyen is funded by the German Ministry of Food and Agriculture (BMEL) Project No. 28DK133A20. }

\author[First]{Maryam Nezami} 
\author[First]{Hossam Seddik Abbas}
\author[Second]{Ngoc Thinh Nguyen} 
\author[First]{Georg Schildbach}
%\author[Forth]{Georg Schildbach}

\address[First]{Institute for Electrical Engineering in Medicine,
        University of Lübeck, Lübeck, Germany (e-mail:
        {\tt\small \{maryam.nezami, h.abbas, georg.schildbach\}@uni-luebeck.de})}
\address[Second]{Institute for Robotics and Cognitive Systems, University of Lübeck, Lübeck, Germany
        (e-mail:{ \tt\small nguyen@rob.uni-luebeck.de})}

\begin{abstract}
This paper proposes a control architecture for autonomous lane keeping by a vehicle. In this paper, 
the vehicle dynamics consist of two parts: lateral %dynamics 
and longitudinal dynamics. Therefore, the control architecture comprises two subsequent controllers. 
A longitudinal \emph{model predictive control} (MPC) makes the vehicle track the desired longitudinal speeds that are assumed to be generated by a speed planner. The longitudinal speeds are then passed to a lateral MPC for lane keeping.
% allows the longitudinal dynamics to follow the desired longitudinal speed for the vehicle, which is .
% For a successful lane keeping, the longitudinal speed should follow the desired reference generated by a speed planner. 
 Due to the dependence of the lateral dynamics on the longitudinal speed, they are represented in a %standard 
\emph{linear parameter-varying} (LPV) form, where its \emph{scheduling parameter} %of the LPV lateral model 
is the longitudinal speed of the vehicle. 
% The lateral dynamics are controlled by an LPV-MPC scheme.
In order to deal with the imprecise information of the future longitudinal speed (the scheduling parameter),
a bound of uncertainty is considered around the nominal trajectory of the future longitudinal velocities.
Then, a tube-based LPV-MPC is adopted to control the lateral dynamics for attaining the lane keeping goal.
% a nominal trajectory of the future longitudinal velocity is considered with a bound of uncertainty around it. 
%which is required over the LPV-MPC prediction horizon, 
%is designed to control the longitudinal velocity of the vehicle. The nominal trajectory of the future longitudinal velocity is considered with a bound of uncertainty around it. Then, a tube-based LPV-MPC is adopted to control the lateral dynamics for attaining the lane keeping goal.
% which can take into account possible additive uncertainties about ({\color{red} add two words here}).
%In the robust LPV-MPC design, the nominal values of the longitudinal speed are used. 
%Due to the Moreover, the state and input constraints are tightened by parameterization of homothetic tubes. 
% Moreover, a method for computing parameterized homothetic tubes with fewer number of vertices is proposed. 
In the end, the effectiveness of the proposed methods is illustrated by carrying out simulation tests. 
\end{abstract}

\begin{keyword}
Automotive Dynamic, Uncertain Systems - LPVS, Predictive Control, Optimal Control, Robustness Issues. 
\end{keyword}

\end{frontmatter}
%%%%%%%%%%%%%%%%%%%%%%%%%%%%%%%%%%%%%%%%%%%%%%
\section{Introduction}
In recent decades, advancements in technology have led to growing attention to autonomous vehicles. An important aspect of autonomous vehicles is their safety. Since vehicles are dynamic systems in a dynamic environment, ensuring their safe performance means assuring their robust performance. 
%Safety of control actions is one aspect of the safe performance of autonomous vehicles.
This means that autonomous vehicles should be able to generate safe control actions when there are changes in the vehicle model or the environment. 
% road traffic, extreme weather conditions, or even internal changes in the vehicle model. 
% Model predictive control (MPC) is a model-based control technique that can guarantee the satisfaction of state and input constraints. In MPC, when solving an online optimization problem, the systems model is exploited to generate the most optimal control input. 

Model predictive control (MPC) is one of the widely used control techniques in autonomous driving in recent years. In some control architectures, MPC is used directly as the controller. For example,
% ~\cite{falcone2007model} proposed an MPC to make the vehicle follow a given path. The proposed MPC is designed for combined steering and braking of the vehicle. 
~\cite{batkovic2020robust} suggested a robust MPC for obstacle avoidance scenarios and a feedback policy to decrease the conservatism of the robust MPC. 
Besides using MPC as a controller, some papers suggest using MPC for providing safe control architectures (SCA) to guarantee the safety of control inputs from unknown resources. For example, \cite{nezami2021safe} proposed a SCA in which uses an MPC as the supervisor for an operating controller and the supervisor intervenes only when necessary. 
% \cite{brudigam2022safe} suggested an architecture that allows the optimal control inputs generated by a stochastic MPC to be applied to the vehicle as long as a safe backup planner can ensure the satisfaction of constraints. 
\cite{tearle2021predictive} designed a predictive safety filter based on MPC. However, in many of the papers exploiting MPC as the controller or supervisor, a common assumption is to consider a constant longitudinal speed for the vehicle, e.g., \cite{nezami2021safe, nezami2022safe, tearle2021predictive,brudigam2021stochastic}.   

% The lateral dynamics and the longitudinal dynamics are two important aspects when modeling a vehicle. 
A vehicle is a nonlinear system; however, it can be formulated as a linear parameter varying (LPV) model~(\cite{hashemi2012low}). The studies on the LPV representation of a vehicle model, e.g.,~\cite{Rajamani2012}, show that the longitudinal speed is a scheduling parameter, which leads to a typical LPV lateral vehicle model. This means that to guarantee the safe application of MPC in vehicles, the internal changes in the speed should be accounted for in the vehicle model. 
%That motivates us to look into the current state of the art in LPV-MPCs, which are very rich. 
% The strength of the LPV-MPC framework is that its optimization problem can still be solved as a \emph{quadratic} or \emph{linear program}, for which efficient solvers are available, and it can yield stability and recursive feasibility guarantees using linear systems tools~(\cite{abbas2019tube}).

One of the main difficulties of LPV-MPC is that the scheduling parameter of the LPV predictor is known only instantaneously. But its future values, which are required over the MPC prediction horizon, are unknown. 
%Therefore, several robust LPV-MPC schemes have been introduced to handle such difficulty. 
One common approach to deal with such difficulty is to accommodate tube-based MPC for the LPV setting, as proposed by~\cite{hanema2016tube}, where
%In~\cite{abbas2016robust}, using a robust MPC for controlling input-output LPV models subject to state and input constraints is suggested. A quadratic terminal cost and an ellipsoidal terminal set are enforced in the optimization problem to guarantee stability. In~\cite{verhoek2021data}, a data-driven predictive control for reference tracking is proposed. In~\cite{hanema2016tube}, 
an anticipate tube MPC algorithm for LPV systems has been introduced. The method takes nominal future values of the scheduling parameter into account with bounded uncertainty around it.
%, which yields a less conservative LPV-MPC scheme. 
\cite{heydari2021robust} suggested designing a tube MPC with cross-section tube parameterization in the presence of disturbances. 
When it comes to the application of LPV systems in autonomous driving, few papers are available. 
\cite{alcala2019lpv} proposed a cascade control design; an LPV-MPC as an external controller and an LPV-LMI-LQR as an internal controller are used for the trajectory tracking of autonomous vehicles. 

\emph{Contributions:} The contributions of this paper are threefold. First, the lateral dynamics of a vehicle are reformulated in a standard LPV form with an additive disturbance in which the longitudinal speed serves as the scheduling parameter. Such formulation allows a low complexity LPV embedding of the lateral dynamics with one scheduling parameter. Second, to control the lateral dynamics of the vehicle, a robust tube-based LPV-MPC setup that handles additive disturbances is proposed. Third, to confirm the control scheme, simulation results are demonstrated.
%The contributions of the paper are as follows:
% \begin{itemize}
%     \item   
%     % % \item proposing a method for the computation of less complex invariant tubes, which reduces the computational burden when solving the MPC optimization problem, 
%     % \item proposing an LPV state feedback controller in which the gain is chosen according to the current state. The controller allows the computation of the robust positively invariant set efficiently.
%     % simple invariant tube, which reduces the computational burden when solving the MPC optimization problem, 
%     \item providing a robust tube-based LPV-MPC setup that handles additive disturbances to control lateral dynamics of the vehicle.  
%     % to control the LPV lateral model of a vehicle for the lane keeping purpose, 
%     % \item providing simulation results to confirm that the above-mentioned LPV state feedback controller helps to reduce the number of vertices of the terminal set,
%     \item demonstrating simulation results to confirm the control scheme.
% \end{itemize} 

\emph{Contents:} In Section~\ref{model}, the longitudinal dynamics and LPV lateral dynamics of the vehicle are represented. Section~\ref{LPV_arch} illustrates the control architecture. Section~\ref{Invareint_set} explains a method for computing a terminal set for the lateral MPC. Finally, in Section~\ref{results}, the implementation of the lateral MPC and the simulation results are discussed.
%then, the simulation results are indicated and discussed. 

\emph{Notations:} The notations $\mathbb{I}$ and $\mathds{1}$ represent the identity matrix and a vector of ones of the proper size, respectively. The convex hull of points $v^1, v^2$ and $v^3$ is denoted by $\textbf{Conv}(v^1,v^2,v^3)$. 
% The predicted value of the state $z_k$ at time $k+i$ based on the available information at time $k$ is denoted by $z_{i|k}$. 
The notation $X \succ 0$ is used to represent the positive definiteness of the matrix $X$. The weighted norm $\| x \|^2_P$ is defined as $\| x \|^2_P = x^\top P x$. Given two sets $\mathcal{Q}_1, \mathcal{Q}_2 \in \mathbb{R}^n$ the Minkowski sum is defined as $\mathcal{Q}_1 \oplus \mathcal{Q}_2 \triangleq \{ q_1 + q_2 | q_1 \in \mathcal{Q}_1, q_2 \in \mathcal{Q}_2  \}$.
%%%%%%%%%%%%%%%%%%%%%%%%%%%%%%%%%%%%%%%%%%%%%%
\section{LPV Dynamic Model of a Vehicle}
\label{model}
\subsection{LPV Model Setup}
Consider the following representation of discrete-time LPV systems subject to additive disturbances:
\begin{equation}\label{e:lpvsys}
x_{k+1}=A(p_k)x_k+Bu_k+w_k,
\end{equation}
where   $u_k\in\mathbb{R}^{n_{\rm u}}$, $x_k\in\mathbb{R}^{n_{\rm x}}$, $p_k\in\mathcal{P}\subset\mathbb{R}^{n_{\rm p}}$ and $w_k\in\mathcal{W} \subset \mathbb{R}^{n_{\rm x}}$, are the system's input, state,  scheduling parameter, and an additive term, respectively,  at a time index $k$. The sets $\mathcal{P}$ and $\mathcal{W}$ are compact sets defined by
\begin{align}
    \mathcal{P}&:=\{p_k\in\mathbb{R}^{n_{\rm p}}\mid p^{\min}\leq p_k\leq p^{\max}\},\label{e:param-space}\\
    \mathcal{W}&:=\{w_k\in\mathbb{R}^{n_{\rm x}}\mid d^{\min}\leq w_k\leq d^{\max}\}. \label{e:dist-space}
\end{align}
It is assumed that $\mathcal{W}$ contains the origin. We also assume that the scheduling parameter $p_k$ and the additive term $w_k$ are known precisely at any time instant $k$; however, their future evolutions might be partially unknown. Moreover, $A(p_k)$ and $B$ are the system matrices with appropriate dimensions, where $A(p_k)$ depends affinely on $p_k$ as
\[
A(p_k)=A_0+\sum_{j=1}^{n_{\rm p}} A_j p_k^{[j]} 
\]
where $p^{[j]}_k$ indicates the $j$th-entry of $p_k$ and $A_j$ are constant known matrices. 

\begin{figure}
    \centering
  \includegraphics[scale=0.9]{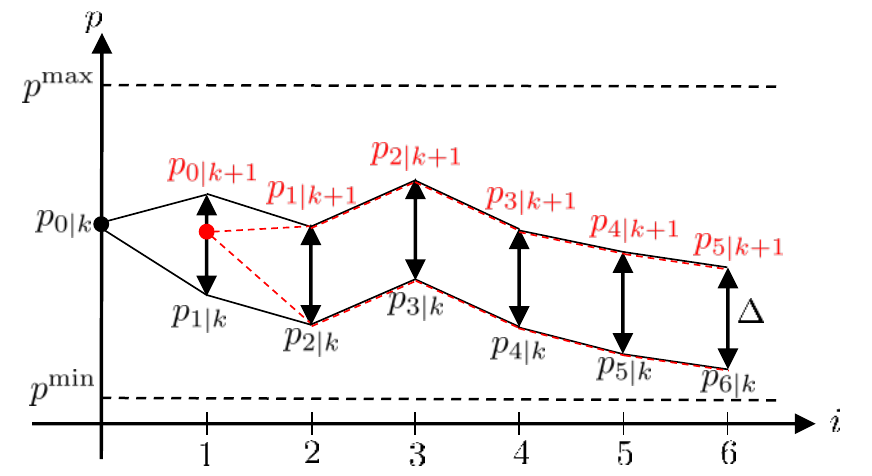}
  \vspace{-1.4em}
  \caption{The illustration of scheduling tubes over the MPC prediction horizon, here $N=6$.}
    \label{fig:scedul_tube}
\end{figure}

For the scheduling parameter, typically, handling the uncertainty in its future values based on the full parameter set $\mathcal{P}$ can lead to a very conservative control scheme.
Therefore, a variety of MPC algorithms propose solutions for this problem, e.g., ~\cite{abbas2018new, hanema2016tube, rakovic2012homothetic}. In this work, we consider an uncertainty bound, i.e., $\Delta$, around such predicted scheduling parameter as follows:
\begin{equation}\label{P_act_value}
    p_{i|k}\in\hat{p}_{i|k} \oplus \Delta, 
\end{equation}
where $\hat{p}_{i|k}$ is the nominal value of the scheduling parameter. This yields a \emph{scheduling tube} of the scheduling parameter trajectory over the MPC prediction horizon 
 as depicted in Fig.~\ref{fig:scedul_tube}, which is more efficient than considering the set $\mathcal{P}$ as the uncertainty set. Then, 
 \begin{equation}
  \underline{p}_{i|k}  \leq p_{i|k}\leq \overline{p}_{i|k},
\end{equation}
$i=1,\cdots,N$, where $\overline{p}_{i|k}$ and $\underline{p}_{i|k}$ represent the bounds associated with 
$\Delta$. An important assumption introduced by~\cite{hanema2016tube} is that the scheduling tube at time instant $k+1$ must be confined inside the tube at the previous time instant. This assumption is required for the recursive feasibility of the proposed MPC. 

Furthermore, we assume that system \eqref{e:lpvsys} is subject to polytopic state and input constraints defined as 
\begin{align}
       \mathbb{X} & :=\{x_k\in\mathbb{R}^{n_{\rm x}} \mid G^{\rm x}x_k\leq h^{\rm x}\}, \label{e:state-contrarian} \\ 
    \mathbb{U} & :=\{u_k\in\mathbb{R}^{n_{\rm u}} \mid G^{\rm u}u_k\leq h^{\rm u}\}. \label{e:input-contrarian} 
\end{align}
where $G^{\rm x} \in \mathbb{R}^{p_{\rm x} \times n_{\rm x} }$, $h^{\rm x} \in \mathbb{R}^{p_{\rm x}}$, $G^{\rm u} \in \mathbb{R}^{p_{\rm u} \times n_{\rm u} }$ and $h^{\rm u} \in \mathbb{R}^{p_{\rm u}}$.

%Since MPC is a model-based control algorithm, this section describes a longitudinal and a lateral vehicle model. 
\subsection{Lateral and Longitudinal Vehicle Dynamics} % LPV Model}
% This section describes the lateral and longitudinal dynamics of the vehicle system used in this work as well as the associated LPV modeling.

\begin{figure}
    \centering
  \includegraphics[scale=0.25]{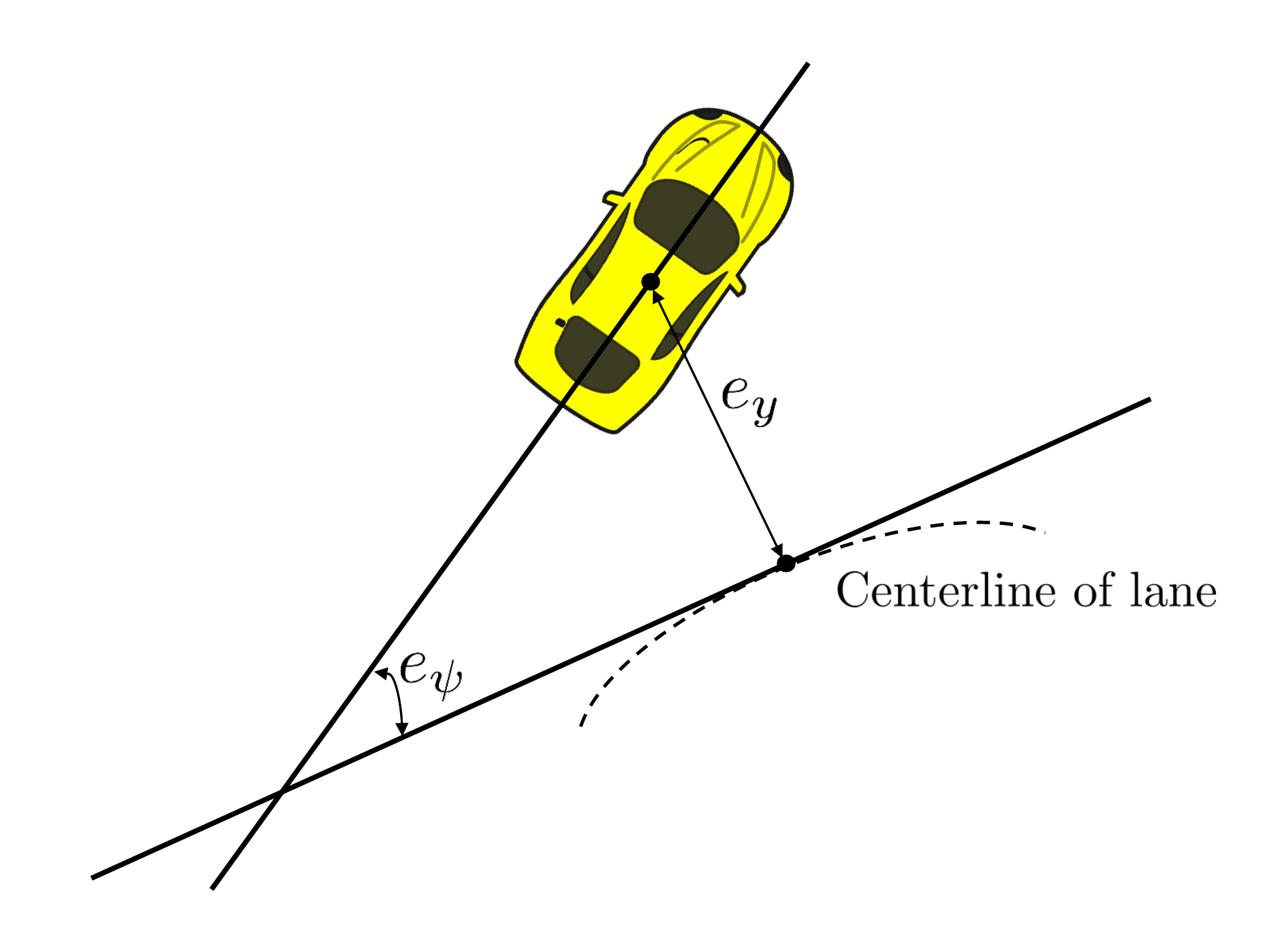}
  \vspace{-2em}
  \caption{Notation of the vehicle~(\cite{carvalho2014stochastic})}
    \label{fig:car_notation}
\end{figure}

%a longitudinal and a lateral vehicle model. 
The continuous time lateral vehicle dynamics are defined using the lateral error model from~\cite[p.~36]{Rajamani2012}:
\begin{multline}
	\!\!\!\!\!\!\!\!\frac{\text{d}}{\text{d}t}\!\!\begin{bmatrix}
		e^{\rm y}(t) \\ \dot{e}^{\rm y}(t)\\ e^{\psi}(t)  \\ \dot{e}^{\psi}(t)
	\end{bmatrix}\!\!\! = \!
\!\!\begin{bmatrix}
			0      &     1    &     0   &   0     \\
			0      &    a    &    b       &   c      \\
			0      &     0    &    0     &   1      \\
			0      &    d     &    e    &    f       \\
	\end{bmatrix}\! \!\!\!\begin{bmatrix}
		e^{\rm y}(t)  \\ \dot{e}^{\rm y}(t)  \\ e^{\psi}(t)  \\ \dot{e}^{\psi}(t) 
	\end{bmatrix}
\!\!	+ \! \!\begin{bmatrix}
		\!	0 \! \\ \! \frac{2C_{\alpha \rm f}}{m} \! \\
		\!	0 \! \\
		\!	\frac{2 C_{\alpha \rm f} l_{\rm f}}{I_{\rm z}} \!\\
	\end{bmatrix}\!\!  \delta(t) \! + f^\psi(t),
	%\!\begin{bmatrix}
%		\!	0 \! \\ \! g \!\\ \! 0 \! \\\!  h \!
%	\end{bmatrix}\!\! \dot{\psi}^{\rm{des}}(t),
	\label{rajamani_model}
\end{multline}
where $e^{\rm y}(t)$ and $e^{\psi} (t)$ are the distance of the center of gravity of the vehicle from the centerline of the lane and the orientation error of the vehicle with respect to the road, respectively, as demonstrated in Fig.~\ref{fig:car_notation}. The states $\dot{e}^{\rm y}(t)$ and $\dot{e}^{\psi}(t)$ are the rate of change of $e^{\rm y}(t)$ and $e^{\psi} (t)$. The control input is the steering angle of the vehicle, i.e., $\delta(t)$. The function $f^\psi(t)$ is defined as  
\begin{equation}\label{e:yaw-rate}
f^\psi(t)= 	\!\begin{bmatrix}	0 & g & 0 &  h 	\end{bmatrix}^\top  \dot{\psi}^{\rm{des}}(t),
\end{equation}
where $\dot{\psi}^{\rm{des}}(t)$ denotes the yaw rate of the road. All the other parameters in \eqref{rajamani_model} and \eqref{e:yaw-rate}
are described as follows:

\begin{minipage}{0.25\textwidth}
\begin{equation*}
	\begin{split}
		a &= -\frac{2C_{\alpha \rm f} + 2 C_{\alpha \rm r}}{{m} v^{\rm x}(t)}, \\
		c &= \frac{-2 C_{\alpha \rm f} l_{\rm f} + 2 C_{\alpha \rm r} l_{\rm r}}{{m} v^{\rm x}(t)},  \\
		e &=  \frac{2 C_{\alpha \rm f} l_{\rm f} - 2 C_{\alpha \rm r} l_{\rm r}}{I_{\rm z}},  \\
		g &= -\frac{2C_{\alpha \rm f}l_{\rm f}-2C_{\alpha \rm r} l_{\rm r}}{{m} v^{\rm x}(t)} \!-\! v^{\rm x}(t),
	\end{split}
\end{equation*}
\vspace{0.008cm}
\end{minipage}
\begin{minipage}{0.2\textwidth}
\begin{equation*}
	\begin{split}
		b &= \frac{2C_{\alpha f} + 2 C_{\alpha \rm r}}{m}, \\
		d &= -\frac{2 C_{\alpha f} l_{\rm f} - 2C_{\alpha \rm r} l_{\rm r}}{I_{\rm z} v^{\rm x}(t)}, \\
		f &=  -\frac{2C_{\alpha f}l^2_{\rm f} + 2 C_{\alpha \rm r} l^2_{\rm r}}{I_{\rm z} v^{\rm x}(t)},  \\
		h &= -\frac{2 C_{\alpha f} l^2_{\rm f} + 2 C_{\alpha \rm r} l^2_{\rm r}}{I_{\rm z} v^{\rm x}(t)},
	\end{split}
\end{equation*}
\vspace{0.008cm}
\end{minipage}

\noindent
 with $v^{\rm x}(t)$ representing  the  longitudinal speed. The constant parameters used in the model, alongside their values and units, are explained in Table~\ref{para-table}.

 %$\dot{\psi}^{\rm{des}}_{k}$ is bounded, i.e., $\dot{\psi}^{\rm{des}}_{k} \in \Psi$ {\color{red} define the set $\Psi$}.  

%Model~\eqref{rajamani_model} is an LPV dynamic model in terms of error with respect to the road with longitudinal speed, $v^{\rm x}(t)$, being the scheduling parameter. As demonstrated in~\cref{fig:car_notation}, $e^{\rm y}(t)$ and $e^{\psi} (t)$ show the distance of the c.g. of the vehicle from the centerline of the lane and the orientation error of the vehicle with respect to the road, respectively. In addition, the rate of change of $e^{\rm y}(t)$ and $e^{\psi} (t)$ are shown with $\dot{e}^{\rm y}(t)$ and $\dot{e}^{\psi}(t)$ in model~\eqref{rajamani_model}. The steering angle of the vehicle is control input and shown with $\delta(t)$. The desired yaw rate of the road is illustrated with $\dot{\psi}^{\rm{des}}(t)$. The parameters used in the model, alongside their values and units, are explained in~\cref{para-table}. 

\begin{table}
	\begin{center}
		\begin{tabular}{ c l c}
			\hline
			\textbf{Symbol} & \textbf{Parameter}& \textbf{Value}  \\
			\hline
			$C_{\alpha \rm f}$ & Cornering Stiffness Front &  $153$\,kN$/$rad \\
			
			$C_{\alpha \rm r}$ & Cornering Stiffness Rear & $191$ \,kN$/$rad \\
			
			$l_{\rm f}$ & Distance CoG to Front Axle & $ 1.3$ \,m \\
			
			$l_{\rm r}$ & Distance CoG to Rear Axle & $1.7$ \,m \\
			
			$I_{\rm z}$ & Vehicle Yaw Inertia & $5250$ \,kgm$^2$ \\
			
			$m$ & Vehicle Mass & $2500$\,kg \\
			
			${\rm l}^{\rm width}$ & lane width &   $10$ $\rm m$ \\
			\hline
		\end{tabular}
	\end{center}
	\caption{Vehicle parameters used in vehicle modeling~(\cite{Gottmann2018})}
	\label{para-table}
\end{table}

Model~\eqref{rajamani_model} is discretized in time using  the Euler discretization rule with a sampling time of $t_s$. Then, the resulted discrete-time model can be realized as an  LPV  model of the form~\eqref{e:lpvsys}, where  $p_k=1/v^{\rm x}_k$, 
%, model~\eqref{rajamani_model} is discretized
%as follows:
%\begin{equation}\label{lat_model}
% 	\bar{x}_{k+1} = A \bar{x}_{k} + B \bar{u}_{k} + E \dot{\psi}^{\rm{des}}_{k}.
%x_{k+1} = A(v^{\rm x}_k) x_{k} + B \delta_{k} + E(v^{\rm x}_k) \dot{\psi}^{\rm{des}}_{k}, 
%\end{equation}
%where 
$x_{k} = \begin{bmatrix} e^{\rm y}_k & \Delta e^{\rm y}_k & e^{\psi}_k & \Delta e^{\psi}_k \end{bmatrix}^\top$, $u_k=\delta_k$, and $w_k=f^\psi_k$, where $k$ is the sampling index. The sets $\mathcal{P}$ and $\mathcal{W}$ can be directly defined by specifying the maximum and minimum values of  $1/v^{\rm x}$ and  $f^\psi$, respectively. It is assumed that $f^\psi(t)$ is bounded with known bounds based on the upper and lower bounds on the road curvature. 
We assume that the exact value of $f^\psi(t)$ is not measurable in advance. Therefore, $f^\psi(t)$ is considered as an additive uncertainty over the MPC prediction horizon, representing  the term $w$ in~\eqref{e:lpvsys}. %\eqref{e:lpvsys}.

The state and input constraints, i.e., $\mathbb{X}$ and $\mathbb{U}$, are defined as \eqref{e:state-contrarian} and \eqref{e:input-contrarian}, respectively,  
%\begin{equation}\label{state_cons_y}
%    \mathbb{X} := \{ x_k \in \mathbb{R}^4 |  G^{\rm x} \hspace{1mm} x_k \leq h^{\rm x} \},
%\end{equation}
where
\begin{align} \label{state_cons_y}
   G^{\rm x} &= \begin{bmatrix} \mathbb{I}_{4 \times 4}  &  -\mathbb{I}_{4 \times 4}  \end{bmatrix}^\top, \qquad    h^{\rm x} = \begin{bmatrix} h & h  \end{bmatrix}^\top,
\end{align}
with $h = \begin{bmatrix} e^{\rm y_{max}} & \Delta e^{\rm y_{max}} & e^{\psi_{\rm max}} & \Delta e^{\psi_{\rm max}} \end{bmatrix}^\top$ and
%The polytopic input constraint formulated is as follows:
%\begin{equation}\label{input_cons_y}
%    \mathbb{U} := \{ \delta_k \in \mathbb{R} | G^{\rm u} \hspace{1mm} \delta_k \leq h^{\rm %\delta}  \}, 
%\end{equation}
%where,
\begin{align} \label{input_cons_y}
    G^{\rm u} &= \begin{bmatrix} 1 & -1  \end{bmatrix}^\top, \qquad h^{\rm u} = \begin{bmatrix} \delta^{\rm max} & \delta^{\rm max}  \end{bmatrix}^\top.
\end{align}

Lane keeping is attained by satisfying the constraint on $e^{\rm y}_k$. In other words, by considering $e^{\rm y_{max}} = \frac{{\rm l}^{\rm width}}{2} - \frac{\rm w^{\rm vehicle}}{2}$, where ${\rm l}^{\rm width}$ is given in Table~\ref{para-table} and $\rm w^{\rm vehicle}$ is the width of the vehicle. Note that $e^{\rm y}_k = 0$ means the vehicle is at the centerline of the lane.

%\subsection{Longitudinal Model}
Next, the continuous longitudinal dynamics of the vehicle are presented as follows:
\begin{equation}\label{long_model}
    \frac{\text{d}}{\text{d}t}\begin{bmatrix}
		s(t) \\ v^{\rm x}(t)
	\end{bmatrix} = 
\begin{bmatrix}
			0      &    1   \\
			0      &    0  
	\end{bmatrix}\! \begin{bmatrix}
		s(t) \\ v^{\rm x}(t)
	\end{bmatrix}
	+ \begin{bmatrix}
			0 \\ 
	     	1  \\
	\end{bmatrix}  a(t),
\end{equation}
where $s(t)$, $v^{\rm x}(t)$, and $a(t)$ denote the longitudinal position of the center of gravity, the longitudinal speed, and the longitudinal acceleration, respectively. 
System~\eqref{long_model} is discretized using the Euler discretization method:
\begin{equation}\label{long_model_disc}
    \begin{bmatrix}
            s_{k+1} \\ v^{\rm x}_{k+1}
    \end{bmatrix} = \begin{bmatrix} 1 &  t_s \\ 0 & 1 \end{bmatrix} \begin{bmatrix}
            s_{k} \\ v^{\rm x}_{k}
    \end{bmatrix} + \begin{bmatrix} 0\\ t_s \end{bmatrix} a_k,
\end{equation}
% \begin{equation}\label{long_model_disc}
%     \tilde{x}_{k+1} = \Tilde{A} \Tilde{x}_{k} + \Tilde{B} a_k,
% \end{equation}
where, $s_k $ and $ v^{\rm x}_k$ are the discretized system states and $t_s$ is the sampling time. The longitudinal speed is bounded as $v^{\rm min} \leq v^{\rm x}_k \leq v^{\rm max}$. The acceleration is bounded as $a^{\rm min} \leq a_k \leq a^{\rm max}$.

% Model~\eqref{long_model_disc} is subject to polytopic state constraint as follows:
% \begin{equation}\label{state_cons_x}
%     \mathcal{X}^{\rm x} := \{ \tilde{x}_k \in \mathbb{R}^2 | \begin{bmatrix} 0 & 1 \\ 0  & -1  \end{bmatrix} \hspace{1mm} \tilde{x}_k \leq  \begin{bmatrix} v^{\rm max} \\ v^{\rm min}  \end{bmatrix} \},
% \end{equation}
% where the values of $v^{\rm max}$ and $v^{\rm min}$ denote the maximum and minimum values of the longitudinal speed.

% System~\eqref{long_model_disc} is subject to polytopic input constraint as follows:
% \begin{equation}\label{input_cons_x}
%     \mathcal{U}^{\rm x} := \{ a_k \in \mathbb{R} | \begin{bmatrix} 1 & 0 \\ 0  & -1  \end{bmatrix}  \hspace{1mm} a_k \leq   \begin{bmatrix} a^{\rm max} \\ a^{\rm min}  \end{bmatrix} \}, 
% \end{equation}
% where the maximum and minimum possible values for the acceleration are denoted by $a^{\rm max}$ and $a^{\rm min}$, respectively.

%%%%%%%%%%%%%%%%%%%%%%%%%%%%%%%%%%%%%%%%%%%%%%
\section{Control Architecture and MPC Design}
\label{LPV_arch}
\begin{figure}
    \centering
  \includegraphics[scale=0.86]{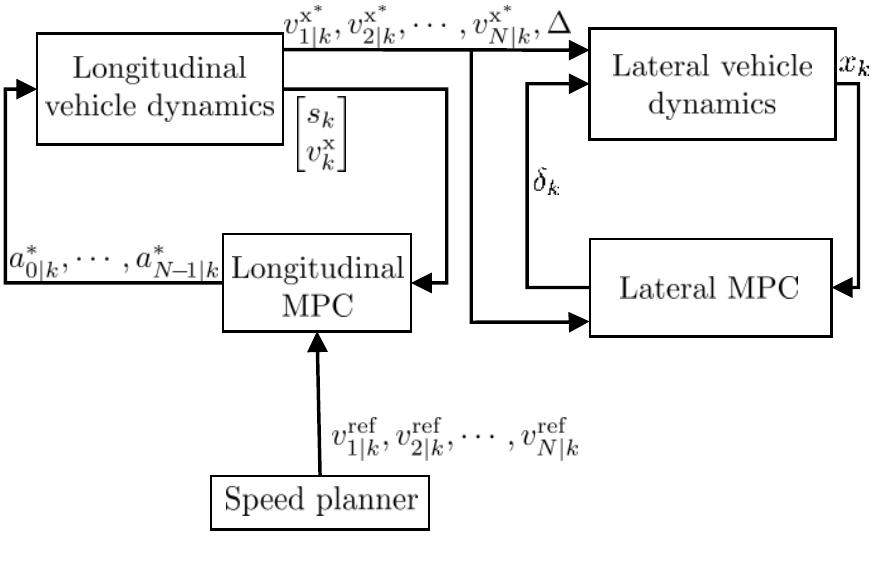}
  \vspace{-2.1em}
  \caption{Depiction of the proposed control architecture}
    \label{fig:control_architecture}
\end{figure}

% As mentioned before, the scheduling parameter of the LPV model of the vehicle %~\eqref{lat_model_dis} 
% is the inverse of the longitudinal speed, i.e., $p_k = 1/v^{\rm x}_k$. 
The proposed control architecture is depicted 
%demonstrated  
in Fig.~\ref{fig:control_architecture}. %As clear from~\cref{fig:control_architecture}, 
A speed planner generates desired reference values for the longitudinal speed of the vehicle. Then, a longitudinal MPC is used to generate the corresponding accelerations to reach the reference speeds while taking %with respect to 
the longitudinal state and input constraints into account. Next, the generated speeds by the longitudinal dynamics are used to schedule the lateral MPC, as explained in the following. 

%\hl{explain the second step of figure}.  

% \begin{figure}
%     \centering
%   \includegraphics[scale=0.9]{Figures/LPV_Arc_V2.pdf}
%   \caption{The control architecture}
%     \label{fig:control_architecture}
% \end{figure}

% \begin{figure}
%     \centering
%   \includegraphics[scale=1]{Figures/LPV_Arc_V3.pdf}
%   \caption{The control architecture}
%     \label{fig:control_architecture}
% \end{figure}

\subsection{Longitudinal MPC}
The following MPC is suggested as the controller for the longitudinal dynamics: 
\begin{align} \label{Longitudinal_MPC}
		\underset{a_{0|k}, \cdots, a_{N-1|k}}{\text{min}}
		&\sum_{i=0}^{N-1} \eta (v^{\rm x}_{i+1|k} - v^{\rm ref}_{i+1|k})^{2} +  \zeta a_{i|k}^2  \nonumber \\
		 \text{s.t.} \;\;
		 & \text{Eq.~\eqref{long_model_disc}}, \quad \forall i = 0,\cdots,N-1, \nonumber \\
% 		&     \begin{bmatrix}
%             s_{i+1|k} \\ v^{\rm x}_{i+1|k}
%     \end{bmatrix} = \begin{bmatrix} 1 &  t_s \\ 0 & 1 \end{bmatrix} \begin{bmatrix}
%             s_{i|k} \\ v^{\rm x}_{i|k}
%     \end{bmatrix} + \begin{bmatrix} 0\\ t_s \end{bmatrix} a_{i|k},   \nonumber \\
		& \begin{bmatrix}
            s_{0|k} & v^{\rm x}_{0|k}
    \end{bmatrix}^\top = \begin{bmatrix}
            s_{k} & v^{\rm x}_{k}
    \end{bmatrix}^\top, \nonumber\\
		&   \begin{bmatrix} 0 & 1 \\ 0  & -1  \end{bmatrix}\! \! \begin{bmatrix}
            s_{i|k} \\ v^{\rm x}_{i|k}
    \end{bmatrix} \leq  \begin{bmatrix} v^{\rm max} \\ -v^{\rm min}  \end{bmatrix}, \nonumber\\
    &
		  \begin{bmatrix} 1  \\ -1  \end{bmatrix}   a_{i|k} \leq   \begin{bmatrix} a^{\rm max} \\ -a^{\rm min}  \end{bmatrix}, 
	\end{align}
where, $\eta > 0$ and $\zeta > 0$ are tuning constants. The current longitudinal state of the vehicle is denoted by $\begin{bmatrix}
            s_{k} & v^{\rm x}_{k}
    \end{bmatrix}^\top$. 

\subsection{Lateral MPC}
The lateral MPC is designed to keep the vehicle in a lane. 
% It is an application of the LPV-MPC scheme as introduced in (1).
% The lateral MPC, to control the lateral vehicle dynamics, is the controller for keeping the vehicle in the lane. 
%  It is an LPV-MPC depending on the vehicle speed, which is related to the scheduling variable, $p_k=1/v^{\rm x}(t_k)$.
 The prediction in such an LPV-MPC scheme is based on \eqref{e:lpvsys}. In this problem, the initial conditions $x_k$, $p_k$ and $w_k$ are available for the MPC at any time instant $k$; however, the prediction of the state over the MPC prediction horizon, i.e., $x_{k+i}$ for $i = 1,\cdots,N$, is a function of $u_{k+i-1}$, $p_{k+i}$ and $w_{k+i}$, which are not available in advance.

 After generation of the accelerations $a^*_{0|k}, \cdots, a^*_{N-1|k}$, by the longitudinal MPC, the future values $v^{\rm x^*}_{1|k},  \cdots, v^{\rm x^*}_{N|k}$ can be computed and used to compute a future scheduling parameter, i.e., $p_{i|k} = 1/v^{\rm x}_{i|k}$,  over the prediction horizon of the lateral MPC.
%  we exploit the information about the future evolution of the vehicle speed generated by the longitudinal MPC to construct a nominal trajectory of the scheduling parameter over the prediction horizon. 
However, due to the physical constraints of the vehicle and inaccuracies of its  model,  
%due to imperfections in the longitudinal model or physical constraints of 
%the vehicle 
the actual future vehicle speed might deviate slightly from what is generated by the longitudinal MPC. In order to take such uncertainty into account, we
employ the available information about  $v^{\rm x^*}_{1|k},  \cdots, v^{\rm x^*}_{N|k}$ to compute nominal values of the future scheduling parameter, i.e., $\hat{p}_{i|k}$, and add a predefined uncertainty region $\Delta$ around such nominal future scheduling parameter, see \eqref{P_act_value}, which leads to a scheduling tube as demonstrated in Fig.~\ref{fig:scedul_tube}. 
% In other words, at every time step $k$, the first value of the scheduling parameter $p_{0|k}=p_k$, which corresponds to the longitudinal speed of the vehicle $v^{\rm x}_{0|k}$, is considered to be exactly known, whereas over the rest of the horizon, the bounded uncertainty $\Delta$ is considered around $\hat{p}_{i|k}$ computed from the related speeds $v^{\rm x^*}_{i|k}$ to take in consideration the uncertainty of ${p}_{i|k}$.
%in other words, it holds that 
In other words, at every time step $k$, the first value of the scheduling parameter $p_{0|k}=p_k$, which corresponds to the longitudinal speed of the vehicle $v^{\rm x}_{0|k}$, is considered to be exactly known. Whereas over the rest of the horizon, the bounded uncertainty $\Delta$ is considered around $\hat{p}_{i|k}$  to take into consideration the uncertainty of ${p}_{i|k}$.
To handle the lateral MPC problem with such scheduling tubes, we adopt a modified formulation of the approach from the paper by~\cite{hanema2016tube} as follows.

%On the other hand, we deal with $w_{k+i}$ as an additive uncertainty over its bounds $\mathcal{W}$. 
 
 %However, as shown in~\cref{fig:control_architecture}, an external disturbance can affect the computed values of $V^{\rm x^*}_{i|k}$.
%Therefore, 
 
%the scheduling parameters will appear as  

 %, and therefore the scheduling parameter in the first step is known.

%. Thus, even though the exact value of the $p_{i|k}$ is not known in every step, its maximum value, i.e., $\overline{p}_{i|k}$, and the minimum value, i.e., $\underline{p}_{i|k}$, are available:
%However, it is important to define the uncertainty bound $\Delta$ in a practical way, as discussed below. 

Consider the following non-empty polytopic set as an invariant set for the LPV system~\eqref{e:lpvsys}:
\begin{equation}\label{polytop_Xn}
    \mathcal{S} = \{ x_k \in \mathbb{R}^{n_{\rm x}} | G^{\rm f} x_k \leq h^{\rm f} \}.
\end{equation}
%In the next section, a method of computing 
Computation of $\mathcal{S}$ can be done offline and will be explained in the next section. A constraint invariant tube using homothetic cross section parameterization, (\cite{rakovic2012homothetic}), is defined as follows: 
\begin{equation} \label{Xk_based_on_X_N}
    X_k := z_k \oplus \alpha_k \mathcal{S}, 
\end{equation}
where $z_k \in \mathbb{R}^{n_{\rm x}}$ and $\alpha_k \in \mathbb{R}_{+}$ represent the center and the radius of the tube, respectively; $z_k, \alpha_k$  are introduced as decision variables in the MPC optimization problem. 
% Then, based on the definition of $\mathcal{S}$ in~\eqref{def_XN}, definition of $X_k$ in \eqref{Xk_based_on_X_N}, can be written as follows:
% \begin{equation}
%     X_k = \textbf{Conv}(\bar{x}^1, \bar{x}^2 , \cdots, \bar{x}^r),
% \end{equation}
% where, $\bar{x}^j = z_j \oplus \alpha_j v^{j}$, $j = 1, \cdots, r$.
Therefore, the lateral MPC is given as  % the LPV model is as follows:
% \begin{align}\label{Lateral_MPC} 
% 		\underset{d_k}{\text{min}}
% 		&  \quad \| X_{N|k} \|^{2}_{P} 
% 		+ \sum_{i=0}^{N-1}  \| X_{i|k} \|^{2}_{Q}  
% 		 + \| u_{i|k} \|^2_{R}  \nonumber \\
% 		 \text{s.t.} \;\;
% 		& \alpha_{0|k} = 0, \quad z_{0|k} = x_k, \nonumber \\
% 		&  \forall i = 0, 1, \cdots, N-1: \nonumber \\
% 		&  X_{i|k} = z_{i|k} \oplus \alpha_{i|k} \mathcal{S} \nonumber, \\
% 		&  u_{i|k} = g_{i|k} \oplus \alpha_{i|k} \mathcal{R},\nonumber \\
% 		& X_{i|k} \in \mathbb{X} \nonumber, \\
% 		& u_{i|k} \in \mathbb{U} \nonumber, \\
% % 		& A(\hat{p}_{i|k} \oplus \Delta) X_{i|k} + B u_{i|k} + (w_{i|k} \oplus \Delta ) \subseteq z_{i+1|k} + \alpha_{i+1|k} \mathcal{S},  \\
% % 		& A(\hat{p}_{i|k} \oplus \Delta) X_{i|k} + B u_{i|k} + (\bar{w}_{i|k} \oplus \Delta ) \subseteq z_{i+1|k} + \alpha_{i+1|k} \mathcal{S},  \\
% 		& A(p_{i|k}) X_{i|k} + B u_{i|k} + (\bar{w}_{i|k} \oplus \Delta_w) \subseteq z_{i+1|k} + \alpha_{i+1|k} \mathcal{S}, \nonumber  \\
% 		& X_{N|k} \in \mathcal{S},
% \end{align}
% where, $p_{i|k} \in \hat{p}_{i|k} \oplus \Delta_p$.
\begin{align}\label{Lateral_MPC} 
		\underset{d_k}{\text{min}}
		&  \quad V_f(z_{N|k},\alpha_{N|k}) 
% 		+ \sum_{i=0}^{N-1}  \| x_{i|k}  \|^{2}_{Q}  
% 		 + \| u_{i|k} \|^2_{R}  \nonumber \\
        + \sum_{i=0}^{N-1}  l(z_{i|k},\alpha_{i|k},g_{i|k}) \nonumber \\
		 \text{s.t.} \;\;
		& \alpha_{0|k} = 0, \quad z_{0|k} = x_k, \nonumber \\
		&  \forall i = 0, 1, \cdots, N-1: \nonumber \\
		&  X_{i|k} = z_{i|k} \oplus \alpha_{i|k} \mathcal{S} \nonumber, \\
		 & u_{i|k} = g_{i|k} \oplus \alpha_{i|k} \mathcal{R},\nonumber \\
% 		& A(\hat{p}_{i|k} \oplus \Delta) X_{i|k} + B u_{i|k} + (w_{i|k} \oplus \Delta ) \subseteq z_{i+1|k} + \alpha_{i+1|k} \mathcal{S},  \\
% 		& A(\hat{p}_{i|k} \oplus \Delta) X_{i|k} + B u_{i|k} + (\bar{w}_{i|k} \oplus \Delta ) \subseteq z_{i+1|k} + \alpha_{i+1|k} \mathcal{S},  \\
		& A(p_{i|k}) X_{i|k} + B u_{i|k} \subseteq z_{i+1|k} + \alpha_{i+1|k} \mathcal{S}, \nonumber  \\
		& X_{i|k} \in \mathbb{X} , \nonumber\\
	   & u_{i|k} \in \mathbb{U} ,\nonumber \\
		&X_{N|k} \in \mathcal{S},
\end{align}

% \begin{align}\label{Lateral_MPC} 
% 		\underset{d_k}{\text{min}}
% 		&  \quad \| X_{N|k} \|^{2}_{P} 
% 		+ \sum_{i=0}^{N-1}  \| X_{i|k} \|^{2}_{Q}  
% 		 + \| u_{i|k} \|^2_{R}  \nonumber \\
% 		 \text{s.t.} \;\;
% 		& \alpha_{0|k} = 0, \quad z_{0|k} = x_k, \nonumber \\
% 		&  \forall i = 0, 1, \cdots, N-1: \nonumber \\
% 		&  X_{i|k} = z_{i|k} \oplus \alpha_{i|k} \mathcal{S} \nonumber, \\
% 		&  u_{i|k} = g_{i|k} \oplus \alpha_{i|k} \mathcal{R},\nonumber \\
% 		& X_{i|k} \in \mathbb{X} \nonumber, \\
% 		& u_{i|k} \in \mathbb{U} \nonumber, \\
% % 		& A(\hat{p}_{i|k} \oplus \Delta) X_{i|k} + B u_{i|k} + (w_{i|k} \oplus \Delta ) \subseteq z_{i+1|k} + \alpha_{i+1|k} \mathcal{S},  \\
% % 		& A(\hat{p}_{i|k} \oplus \Delta) X_{i|k} + B u_{i|k} + (\bar{w}_{i|k} \oplus \Delta ) \subseteq z_{i+1|k} + \alpha_{i+1|k} \mathcal{S},  \\
% 		& A(p_{i|k}) X_{i|k} + B u_{i|k} \subseteq z_{i+1|k} + \alpha_{i+1|k} \mathcal{S}, \nonumber  \\
% 		& X_{N|k} \in \mathcal{S},
% \end{align}
where, $p_{i|k}\in\hat{p}_{i|k} \oplus \Delta$. The decision variable is $d_k = \begin{bmatrix} \alpha_{0|k}, \cdots, \alpha_{N|k}, z_{0|k}, \cdots, z_{N|k}, g_{0|k}, \cdots,g_{N-1|k} \end{bmatrix}^\top$.The input
and state constraints $\mathbb{U}$ and $\mathbb{X}$ are defined in~\eqref{input_cons_y} and~\eqref{state_cons_y}, respectively. The invariant set $\mathcal{S}$ as defined in~\eqref{polytop_Xn} is used
as the terminal set of the MPC and to parameterize the state tube to take into account the uncertainty of $p$ over $N$ and possible additive uncertainty, i.e., $w$, for the system \eqref{e:lpvsys}. 
% and $v^j$ are the vertices of $\mathcal{S}$ as defined in~\eqref{def_XN}. The matrices $G^{\rm f}$, $h^{\rm f}$, $G^{\rm x}$, $h^{\rm x}$, $G^{\rm u}$ and $h^{\rm u}$ are from~\eqref{polytop_Xn},~\eqref{state_cons_y} and~\eqref{input_cons_y}, respectively. 
% The terminal cost is computed as 
% \begin{equation} \label{terminal_cost}
% P_{N|k} = P_0 + P_1 p_{N|k},
% \end{equation}
% where the computation of $P_0$ and $P_1$ is explained in the next section. The gain $K_{i|k}$ is computed as follows:
% \begin{equation} \label{K_gain}
%     K_{i|k} = K_0 + K_1 p_{i|k},
% \end{equation}
% where the computation of $K_0$ and $K_1$ are explained in the next section. 
In the optimization problem~\eqref{Lateral_MPC}, the set $\mathcal{R}$ is used to parameterize the control tube. The simplest way to compute $\mathcal{R}$ is $\mathcal{R} = K\mathcal{S}$, where $K$ 
%can be computed with a constant gain $K$ as $\mathcal{R} = K\mathcal{S}$. The gain $K$ 
is chosen such that $A(p_k) + B K$, is stable for all $p_k \in \mathcal{P}$. %, where $A(p_k)$ and $B$ are from~\eqref{e:lpvsys}. 
However, the constant  $K$  can lead to a very conservative solution for the LPV-MPC~\eqref{Lateral_MPC}.
Alternatively, choosing a parameter-dependent gain $K(p_k)$ is preferred; see next section for more details. In this case, $\mathcal{R}$ in~\eqref{Lateral_MPC} also depends on time and can be computed as follows:
\begin{equation}\label{eq:R}
    \mathcal{R}_{i|k} = K(p_{i|k}) \mathcal{S}, 
\end{equation}
for all $p_{i|k}\in\hat{p}_{i|k} \oplus \Delta$. The implementation of such formulation is explained in the next section. 
% Finally,    
% %Similar to what was just discussed for the computation of $\mathcal{R}$,
% the terminal cost $P$ as used in the optimization problem~\eqref{Lateral_MPC} is a constant positive definite matrix $P=P^\top\succ 0$ satisfying the condition $(A(p)+BK)^\top P (A(p)+BK) - P \leq -({Q}+K^\top {R}K)$, for all $p\in\mathcal{P}$.  
 The cost function in the optimization problem~\eqref{Lateral_MPC} is still in terms of the uncertain state. A worst-case stage cost, based on~\cite{hanema2016tube}, can be computed as
\begin{equation}\label{cost}
    l(z_{i|k},\alpha_{i|k},g_{i|k}) = \underset{(x,u) \in X_{i|k} \times \mathcal{R}}{\text{max}} \| x  \|^2_Q + \| u \|^2_R, 
\end{equation}
where $Q = {Q}^\top  \succ 0 $ and $R = {R}^\top \succ 0$ are the tuning matrices. The terminal cost in~\eqref{Lateral_MPC} can be computed as
\begin{equation}\label{terminal_Cost}
    V_f(z_{N|k},\alpha_{N|k}) = \underset{x \in X_{N|k} }{\text{max}} \| x  \|^2_P,
\end{equation}
where the terminal cost $P$ is a constant positive definite matrix $P=P^\top\succ 0$ satisfying the condition $(A(p)+BK)^\top P (A(p)+BK) - P \leq -({Q}+K^\top {R}K)$, for all $p\in\mathcal{P}$. Similar to what was just discussed for the computation of $\mathcal{R}$, a less conservative terminal cost is to consider $P$
parameter-dependent, i.e., $P(p)=P^\top(p)\succ 0$, $\forall p \in \mathcal{P}$ as explained in the next section.
The main difference between the  LPV-MPC scheme presented here and that in~\cite{hanema2016tube} is that the latter did not consider the additive term, i.e.,
$w_k=0,  \forall k$. Also the paper considers contractive invariant sets. For this reason, the terminal cost in~\eqref{terminal_Cost} is different from that in~\cite{hanema2016tube}. For the stability of the control law generated by this LPV-MPC scheme and its recursive feasibility without additive disturbance, the reader can refer to~\cite{hanema2016tube}, whereas these theoretical results cannot be guaranteed in the presence of additive disturbance. However, any infeasibility or instability have not been observed in the simulations. 

% \hl{
% \textit{Remark}: We do not have concrete proof for the recursive feasibility yet. But, in the simulations, we did not observe any infeasiliby problem.}

% textwidth in cm: \printinunitsof{cm}\prntlen{\textwidth}

% textwidth in cm: \printinunitsof{cm}\prntlen{\linewidth}
%%%%%%%%%%%%%%%%%%%%%%%%%%%%%%%%%%%%%%%%%%%%%%
\section{Computation of the Invariant Set} % Terminal set}
\label{Invareint_set}
% The first step in the computation of $\mathcal{S}$ is to design a state feedback controller for the LPV system \eqref{e:lpvsys}. The optimization problem~\eqref{Lateral_MPC} can be computationally expensive if $\mathcal{S}$ has many vertices. 

% The affine representations of the system matrices in~\eqref{e:lpvsys} are as follows:
% \begin{equation} \label{aff_rep}
%     A(p_k) = A_0 + p_k A_1, \quad B = B_0 + p_k B_1, 
% \end{equation}
% where, $A_0$, $A_1$ and $B_0$ are constant matrices. Since $B$ does not depend on $p_k$, $B_1$ is a matrix of zeros. 

Consider the LPV system~\eqref{e:lpvsys}, where $p_k \in \mathcal{P}$ is presented in~\eqref{e:param-space}. Using an LPV state feedback controller as follows
\begin{equation}\label{control_law}
    u_k = K(p_k)x_k,
\end{equation}
system~\eqref{e:lpvsys} is called \emph{poly-quadratically stabilized} 
according to~\cite{pandey2017quadratic} if the \emph{linear matrix inequality} (LMI) conditions shown below are satisfied. 
% In \eqref{control_law}, the affine representation of $K(p_k)$ is as follows: 
% \begin{equation}\label{lpv_k}
%     K(p_k) = K_0 + p_k K_1,  
% \end{equation}
% where, $K_0 \in \mathbb{R}^{1 \times 2} \text{ and } K_1 \in \mathbb{R}^{1 \times 2}$. 
 
Consider the   vertex representation of the system matrix of \eqref{e:lpvsys} as
$A(p_k) =  \textbf{Conv}(A^1, A^2,\cdots, A^{n_{\rm v}})$ %and $B = \textbf{Conv}(B^1,B^2)$ to be the vertex representation of $A(p_k)$ and $B$ from~\eqref{e:lpvsys}, 
where $A^1$, $A^2$, $\cdots$ are the evaluation of $A(p_k)$ at the $n_{\rm v}$ vertices of the set $\mathcal{P}$.
%$B^1$ and $B^2$ to show the vertices of $A(p_k)$ and $B$ and are computed as follows
%\begin{equation}
%    A^1 = A(p^{\rm min}), \quad A^2 = A(p^{\rm max}),  \quad B^1= B^2 = B.
%\end{equation}
Then, %the vertices of 
$K(p_k)$ can be computed by solving the following set of LMIs~(\cite{bao2022learning}):
% \begin{equation}
%     \begin{bmatrix}
%             X^j \!+ \!{X^j}^\top \!- \!S^j & *  & * & * & * \\
%             {A^j}^\top X^j & S^l-T^{j,l} & * & * & * \\
%             -W^j & {Z^l}^\top{B^j}^\top -Y^l & Z^l + {Z^l}^\top & * & * \\ 
%             Q^{1/2}X^j & 0 & 0 & I & * \\
%             R^{1/2}W^j & 0 & 0 & 0 & I
%     \end{bmatrix} \succ 0
% \end{equation}
\begin{equation} \label{LMI}
    \begin{bmatrix}
            X^j \!\!+ \!\!{X^j}^\top \!\!\!- \!S^j & *  & * & * & * \\
            {A^j}^\top X^j & S^l\!-\!B Y^l \!+ \!{Y^l}^\top \! {B}^\top & * & * & * \\
            -W^j & {Z^l}^\top{B}^\top -Y^l & Z^l\! \!+ \!\!{Z^l}^\top & * & * \\ 
            {Q}^{1/2}X^j & 0 & 0 & \mathbb{I} & * \\
            {R}^{1/2}W^j & 0 & 0 & 0 & \mathbb{I}
    \end{bmatrix} \!\succ \!0, 
\end{equation}
for $j,l = 1,2,\cdots, n_{\rm v}$ such that the matrices $X^j, W^j, Y^j$ and $S^j \succ 0$. 
Thus, the feedback gain is computed as $K^{j} = W^j{X^j}^{-1}$, where $K^{j}$ denotes the $j^{\rm th}$-vertex according to  $K(p_k) = \textbf{Conv}(K^1,K^2,\cdots,K^{n_{\rm v}})$.
% Having the vertex representation of $K^{j}$, $K_0$ and $K_1$ for the affine representation can be computed and then be used in~\eqref{K_gain}.
This yields the associated Lyapunov matrix $P$ parameter-dependent such that $P(p_k) = \textbf{Conv}(P^1,P^2,\cdots, P^{n_{\rm v}})$ where $P^j = {S^j}^{-1}$.
%The vertices of the matrix $P$ can be computed as $P^j = {S^j}^{-1}$, then 
% it can be transformed to the affine representation by computing $P_0$ and $P_1$ and to use it the terminal cost in the optimization problem~\eqref{Lateral_MPC} as shown in~\eqref{terminal_cost}. 

The usage of $P(p_k)$ and $K(p_k)$ after their computation by solving the LMI problem~\eqref{LMI} is indicated in the next section.
% where the implementation of the optimization problem~\eqref{Lateral_MPC} is explained. 

%Next, consider the following definition:
\begin{definition}[\textbf{Robust positive invariant set}]\label{d:RPI}
		\normalfont{The \\
		set $\mathcal{S} \in \mathbb{X} $ is said to be a \textit{robust positively invariant (RPI) set} for the system~\eqref{e:lpvsys} with $u_k = K(p_k)x_k \in \mathbb{U}$, if for all $ p_{k} \in \mathcal{P}$ and all $w_k \in \mathcal{W}$, having $x_{k} \in \mathcal{S}$ leads to $ x_{k+1} \in \mathcal{S}$~(\cite{Blanchini1999}).}
\end{definition}

For the LPV lateral vehicle model, represented in the form~\eqref{e:lpvsys}, possible changes in the curvature of the road are considered as additive disturbances, which are considered in the computation of  $\mathcal{S}$ given the set $\mathcal{W}$. Therefore, anytime the road profile changes, and consequently, the upper and lower bounds on the road curvature change, it means that the disturbance set has changed, then $\mathcal{S}$ has to be recomputed. 
% Therefore, the controller should be updated anytime the road profile, and consequently, the upper and lower bounds on the road curvature change.
Computation of  $\mathcal{S}$ is performed offline. 

%as an  RPI set according to Definition~\ref{d:RPI}. 

%even though its values are available over the prediction horizon. 
%The reason is that in order to reduce the computational time, online computation of $\mathcal{S}$ is not desirable. But, to compute an $\mathcal{S}$ that can be used all over the road, based on the information that we have from the road at the beginning. If the road changes, and consequently the road curvature changes, a new $\mathcal{S}$ has to be computed. There are many papers on the computation of 

% In this paper, the algorithm proposed by~\cite{Nguyen2014} is used for the computation of RPI sets as in Definition~\ref{d:RPI}. 

% {\color{red} you should also write a few words about the commutation of $\mathcal{R}$ when parameter dependent $K$ is used.}

% {\color{red} complete this section by showing the algorithm to compute  S and motivate why you need to consider the additive uncertainty}
%%%%%%%%%%%%%%%%%%%%%%%%%%%%%%%%%%%%%%%%%%%%%%

\section{Results and Discussions}
\label{results}
\subsection{Implementation}
% First, the method of implementing the
% optimization~\eqref{Lateral_MPC} is explained. 
Let the set $\mathcal{S}$ be presented as:
\begin{equation}\label{def_XN}
    \mathcal{S} = \textbf{Conv}(v^1,v^2 \cdots, v^{r}),
\end{equation}
where $v^1,v^2 \cdots, v^{r}$ are the vertices of $\mathcal{S}$. 
%Assuming $v^1,v^2 \cdots, v^{r}$ to be the vertices of $\mathcal{S}$, it can be presented as follows: 
Based on the vertices representation of $\mathcal{S}$, the optimization problem~\eqref{Lateral_MPC} can be implemented as follows:
\begin{align}\label{Lateral_MPC_imple} 
		\underset{d_k}{\text{min}}
		& \sum_{j=1}^{r}  \| z_{N|k} \oplus \alpha_{N|k} v^{j} \|^{2}_{P^l} \nonumber \\
		&+ \sum_{i=0}^{N-1} \sum_{j=1}^{r}  \| z_{i|k} \oplus \alpha_{i|k} v^{j} \|^{2}_{Q}  
		 + \| g_{i|k} + \alpha_{i|k} K^l v^{j} \|^2_{R}  \nonumber \\
		 \text{s.t.} \;\;
		& \alpha_{0|k} = 0, \quad z_{0|k} = x_k, \nonumber \\
		&  G^{\rm f}( z_{N|k} \oplus \alpha_{N|k} v^{j} )\leq h^{\rm f} \nonumber, \\
		&  G^{\rm f}( A(\overline{p}_{i|k})(z_{i|k} \!\oplus \!\alpha_{i|k} v^{j}\!) \! + \! B (g_{i|k}\! + \! \alpha_{i|k} K^l v^{j}) ) \! \nonumber \\
		& \hspace{44 mm}\leq \! G^{\rm f} z_{i+1|k} \! + \! \alpha_{i+1|k} h^{\rm f} \! \nonumber, \\
		&  G^{\rm f} ( A(\underline{p}_{i|k} )(z_{i|k} \!\oplus \!\alpha_{i|k} v^{j}\!) \! + \! B (g_{i|k} \!+\! \alpha_{i|k} K^l v^{j}) ) \! \nonumber \\
		& \hspace{44 mm}\leq \! G^{\rm f} z_{i+1|k} \! + \! \alpha_{i+1|k} h^{\rm f} \! \nonumber, \\
% 		&  G^{\rm f}( A(\underline{p}_{i|k} )(z_{i|k} \oplus \alpha_{i|k} v^{j}) + B u_{i|k} )\leq  G^{\rm f} z_{i+1|k} + \alpha_{i+1|k} h^{\rm f} \nonumber, \\
		& G^{\rm x} ( A(\overline{p}_{i|k} )(z_{i|k} \!\oplus\! \alpha_{i|k} v^{j}) \!+ \!B (g_{i|k} \! + \! \alpha_{i|k} K^l v^{j}) )  \leq h^{\rm x},  \nonumber \\
% 		& \hspace{70mm}\leq h^{\rm x} , \nonumber\\
		& G^{\rm x} ( A(\underline{p}_{i|k} )(z_{i|k} \! \oplus \! \alpha_{i|k} v^{j}) \!+ \!B (g_{i|k} \!+ \! \alpha_{i|k} K^l v^{j})) \leq h^{\rm x}, \nonumber  \\
% 		& \hspace{70mm}\leq h^{\rm x} , \nonumber\\
		& G^{\rm u} (g_{i|k} + \alpha_{i|k} K^l v^{j}) \leq h^{\rm u}, \\
		 & \forall i = 0, 1, \cdots, N\!-\!1, \quad \forall j = 1, 2,\cdots,r, \quad \forall l =1,2, \nonumber 
	\end{align}
where $v^{j}$ are the vertices of $\mathcal{S}$ from~\eqref{def_XN},
$j,l$ denote the vertex number of the sets  $\mathcal{S}$ and  $\mathcal{P}$, respectively.
The parameters $K^l$ and $P^l$ are the vertices of LPV state feedback gain and the LPV terminal weight, calculated by solving the LMI problem~\eqref{LMI}. Note that, with one scheduling parameter for the LPV lateral model, the number of vertices of $\mathcal{P}$ is $n_{\rm v}=2$.
The rest of the parameters are the same as for the optimization problem~\eqref{Lateral_MPC}. 

\subsection{Simulation Results}
% Observing optimization problem~\eqref{Lateral_MPC_imple}, it is clear that its computation time highly depends on the number of vertices of $\mathcal{S}$. A comparison of the number of vertices of $\mathcal{S}$, between using a constant gain $K$ and a parameter-dependent gain $K(p_k)$ as in~\eqref{control_law} is illustrated in Table~\ref{table_com_s}. By choosing different values for $Q$ and $R$ number of vertices of $\mathcal{S}$ significantly changes, while with a constant gain, it is not affected. The results confirm that using LPV feedback gain can significantly reduce the number of $\mathcal{S}$ vertices and computation time. However, it is worth mentioning that even though the number of vertices of $\mathcal{S}$ is reduced by using the LPV state feedback, its volume is also decreasing.

% \hl{The simulation is implemented on} a Dell Latitude 5590 with windows 10, Intel(R) Core(TM) i7-8650U CPU, and 16 GB RAM. 

\begin{figure}
    \centering
  \includegraphics[scale=1]{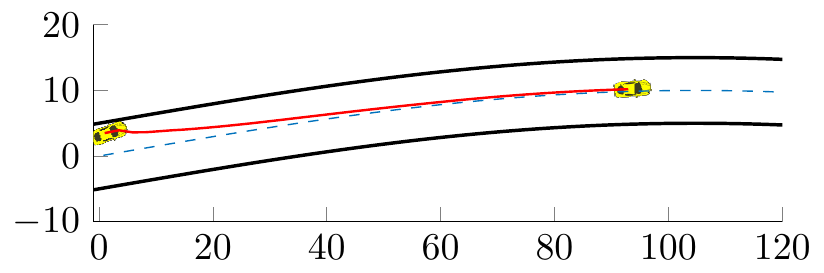}
 \vspace{-1.0em}
  \caption{A sample lane keeping scenario }
    \label{fig:lane}
\end{figure}

An example of a lane keeping scenario by the proposed control architecture is illustrated in Fig.~\ref{fig:lane}. The scenario is simulated using YALMIP (\cite{Lofberg2004}) and solved by quadprog solver in~\cite{MATLAB:2019}. In this scenario, the vehicle is controlled to approach the centerline of the road using the proposed control architecture. The parameters used in the lateral and longitudinal MPC in this scenario are given in Table~\ref{table_mpc}. The initial conditions are chosen as $e^{\rm y}_0 = 3.27$ $\rm m$, $\dot{e}^{\rm y}_0 = 0.55$ $\frac{\rm m}{\rm s}$, $e^{\psi}_0 = -0.24$ $\rm rad$, $\dot{e}^{\psi}_0 = 0.3$ $\frac{\rm rad}{\rm s}$, $s_0 = 1$ $\rm m$ and $v^{\rm x}_0 = 25 $~$\frac{\rm m}{\rm s}$. 
% The simulation was implemented on a Dell Latitude 5590 with windows 10, Intel(R) Core(TM) i7-8650U CPU, and 16 GB RAM. However, to check the real-time applicability of the control approach stronger processor is needed.

For the computation of $\mathcal{S}$, at first, an LPV state feedback controller, as explained in Section~\ref{Invareint_set}, is designed. 
It has been observed that changing $Q$ and $R$ highly affects the number of vertices of $\mathcal{S}$, which has a great impact on the number of inequality constraints in the related MPC problem. Similar to the method presented by~\cite{nguyen2020stability}, different values for $Q$ and $R$ have been tried. Among the test values, the values for $Q$ and $R$, which result in the least number of vertices for $\mathcal{S}$, have been selected as the final values. The final selected values for $Q$ and $R$ can be seen in Table~\ref{table_mpc}. The set $\mathcal{S}$ is computed by using the algorithm for RPI set computation by~\cite[p.~25]{Nguyen2014}. This choice results in $93$ vertices for $\mathcal{S}$, which leads to 1770 inequality constraints when setting up the MPC.

\begin{table}
	\begin{center}
		\begin{tabular}{ c l | c l}
			\hline
			\textbf{Parameter} & \textbf{Value}&  \textbf{Parameter} & \textbf{Value}  \\
			\hline
			$\eta$ & $100$ &  $\zeta$ & $0.1$ \\
			$a_{\rm min}$ & $-6\frac{\rm m}{\rm s^2}$ & $a_{\rm max}$ & $2\frac{\rm m}{\rm s^2}$\\
			$N$ & $5$ & $t_s$ & $0.1$ s\\
			$e^{\rm y_{max}}$ & $4$m & $\Delta e^{\rm y_{\rm max}} $	& $10\frac{\rm m}{\rm s}$ \\
			$e^{\psi_{\rm max}}$   & $\frac{\rm pi}{2}$ rad & $\Delta e^{\psi_{\rm max}}$ & 
			$\frac{\rm pi}{3 t_s}$ $\frac{\rm rad}{ \rm s}$ \\
			$v^{\rm max}$ &   $30 \frac{\rm m}{\rm s}$  & $v^{\rm min}$ &  $15\frac{\rm m}{\rm s}$ \\
			$Q$ & $50 \mathbb{I}_{4 \times 4}$ & $R$ &  $5$ \\
			$\Delta$ & $0.2$ & $d^{\max} = - d^{\min} $ & $10^{-2} \mathds{1}_{4 \times 1}$ \\
			\hline
		\end{tabular}
	\end{center}
	\caption{The MPC parameters}
	\label{table_mpc}
\end{table}

\begin{figure}
    \centering
  \includegraphics[scale=1]{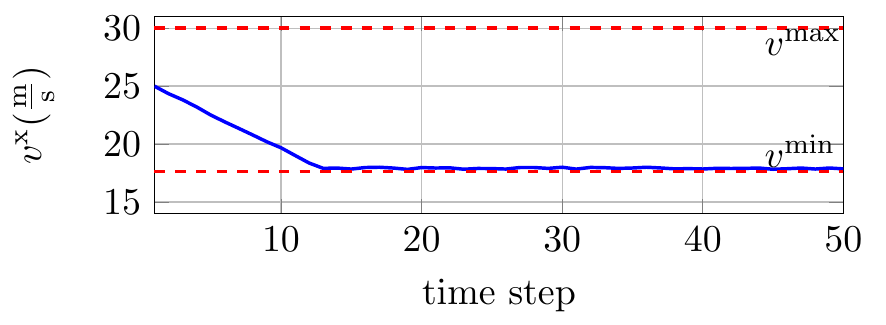}
  \vspace{-1.9em}
  \caption{Longitudinal speed of the vehicle}
    \label{fig:speed}
\end{figure}

\begin{figure}
    \centering
  \includegraphics[scale=1]{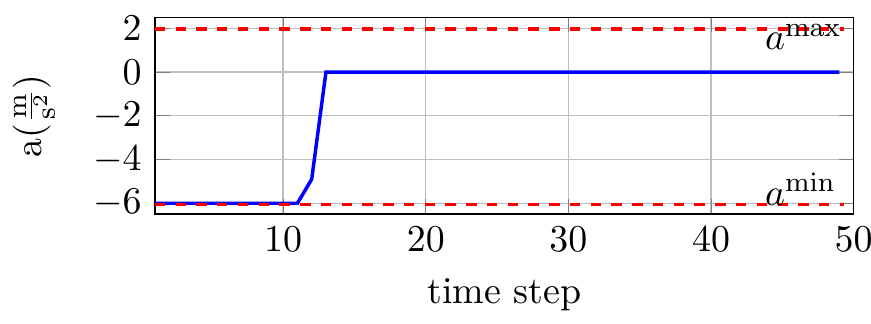}
  \vspace{-2.4em}
  \caption{Longitudinal acceleration of the vehicle}
    \label{fig:acc}
\end{figure}

\begin{figure}
    \centering
  \includegraphics[scale=1]{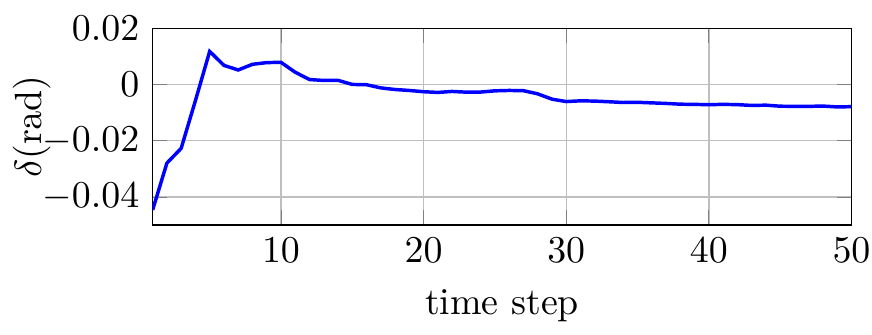}
  \vspace{-3em}
  \caption{Steering angle}
    \label{fig:steer}
\end{figure}

\begin{figure}
    \centering
  \includegraphics[scale=1]{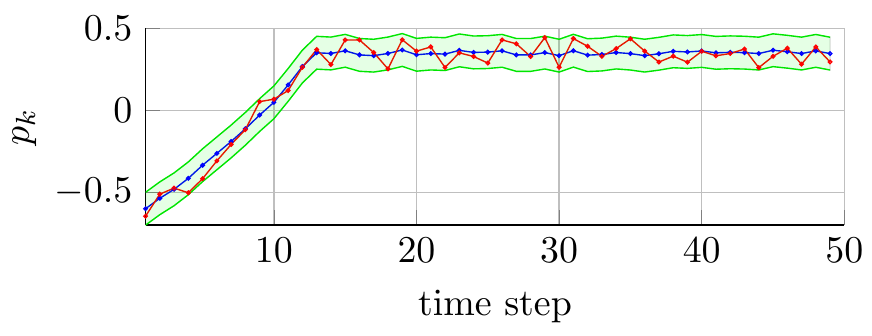}
  \vspace{-2.4em}
  \caption{The scheduling parameter }
    \label{fig:sch}
\end{figure}

The reference speed generated by the speed planner is assumed to be $18$ $\frac{\rm m}{\rm s}$. As illustrated in Figs.~\ref{fig:speed} and \ref{fig:acc}, the vehicle decelerates with the maximum value until its longitudinal speed reaches the reference value of $18$ $\frac{\rm m}{\rm s}$. The steering angles generated by the lateral MPC to bring the vehicle to the centerline of the road are demonstrated in Fig~\ref{fig:steer}. In Fig.~\ref{fig:sch}, the blue line indicates the nominal values of the scheduling parameter, i.e., $\hat{p}_k$, and the red line represents the actual values, i.e., $p_k$. The green tube around the blue line is plotted based on the value of $\Delta$, given in Table~\ref{table_mpc}. In the simulations, the deviation of $p_k$ from its nominal value, based on the value of $\Delta$, is generated randomly. 

As the sample scenario confirms, by using the proposed architecture, the vehicle can track the reference, %the centerline of the vehicle, 
while the state and input constraints are satisfied. Furthermore, the simulation results confirm that the lateral MPC is robust against changes in the scheduling parameter.
% In this scenario, the speed of the vehicle at the moment that simulation starts is $25 \frac{\rm m}{\rm s}$. The speed planner generates a constant speed of 

%{\color{red} this section is not complete}

% In this section, in two scenarios, the applicability of the proposed control architecture is showcased. The parameters that are used in the simulation of the vehicle model~\eqref{rajamani_model} are listed in~\cref{para-table}. 

\section{Conclusions}
\label{conclusion}
In this paper, the lateral dynamics of a vehicle with an additive term have been embedded in an LPV representation, where the speed of the vehicle is the scheduling parameter. A robust control architecture consisting of two subsequent controllers is proposed for lane keeping. At first, a longitudinal MPC for controlling the longitudinal dynamics has been employed. Then, the longitudinal MPC provides an estimate of the vehicle speed to be used by the subsequent MPC for controlling the lateral dynamics. For robustification of the lateral MPC, an uncertainty region around the estimated speeds (the scheduling parameter) has been considered. Still, there is a lot to be explored in this area. e.g., extending the work to obstacle avoidance scenarios or providing guarantees for stability and recursive feasibility.

% This paper proposes a robust control architecture consisting of two subsequent controllers for lane keeping. At first, a longitudinal MPC for controlling the longitudinal dynamics has been employed. Next, the resulted longitudinal speeds from the longitudinal MPC with an uncertainty region around them have been considered as a scheduling parameter to re-formulate the lateral dynamics of the vehicle in a standard LPV form. Then, a robust tube-based LPV-MPC has been utilized to control the LPV lateral dynamics. 
% Still, there is a lot to be explored in this area. e.g., extending the work to obstacle avoidance scenarios or providing guarantees for stability and recursive feasibility.

% it is especially interesting to look at scenarios where the longitudinal speed of the vehicle is changing drastically and compare the scenarios with that of a conventional MPC.

\bibliography{ifacconf}             % bib file to produce the bibliography
                                                     % with bibtex (preferred)
                                                   
%\begin{thebibliography}{xx}  % you can also add the bibliography by hand

%\bibitem[Able(1956)]{Abl:56}
%B.C. Able.
%\newblock Nucleic acid content of microscope.
%\newblock \emph{Nature}, 135:\penalty0 7--9, 1956.

%\bibitem[Able et~al.(1954)Able, Tagg, and Rush]{AbTaRu:54}
%B.C. Able, R.A. Tagg, and M.~Rush.
%\newblock Enzyme-catalyzed cellular transanimations.
%\newblock In A.F. Round, editor, \emph{Advances in Enzymology}, volume~2, pages
%  125--247. Academic Press, New York, 3rd edition, 1954.

%\bibitem[Keohane(1958)]{Keo:58}
%\newblock \emph{Power and Interdependence: World Politics in Transitions}.
%\newblock Little, Brown \& Co., Boston, 1958.

%\bibitem[Powers(1985)]{Pow:85}
%\newblock Is there a way out?
%\newblock \emph{Harpers}, pages 35--47, June 1985.

%\bibitem[Soukhanov(1992)]{Heritage:92}
%\newblock \emph{{The American Heritage. Dictionary of the American Language}}.
%\newblock Houghton Mifflin Company, 1992.

%\end{thebibliography}

\appendix

\end{document}